\documentclass[apjl]{emulateapj}
\usepackage{epstopdf}
\usepackage{natbib}
\usepackage{amsmath}
\usepackage{enumerate}
\def\simpropto{\lower.2ex\hbox{$\; \buildrel \propto \over \sim \;$}}
\def\ltsim{\lower.5ex\hbox{$\; \buildrel < \over \sim \;$}}
\def\gtsim{\lower.5ex\hbox{$\; \buildrel > \over \sim \;$}}

\usepackage{color}

\begin{document}

\title{Ultrahigh Energy Cosmic Rays and Black Hole Mergers}

\author{Kumiko Kotera\altaffilmark{1,2}}

\author{Joseph Silk\altaffilmark{1,2,3,4}}
\altaffiltext{1}{Institut d'Astrophysique de Paris, UMR 7095, CNRS, UPMC Universit\'{e} Paris 6, Sorbonne Universit\'{e}s, 98 bis boulevard Arago, 75014 Paris, France}
\altaffiltext{2}{Laboratoire AIM-Paris-Saclay, CEA/DSM/IRFU, CNRS, Universite Paris Diderot,  F-91191 Gif-sur-Yvette, France}

\altaffiltext{3} {Department of  Physics \& Astronomy, The Johns Hopkins University, 
3400 N Charles Street, Baltimore, MD 21218, USA}\altaffiltext{4} {Beecroft Institute of Particle Astrophysics and Cosmology, Department of Physics,
University of Oxford, Denys Wilkinson Building, 1 Keble Road, Oxford OX1 3RH, UK}

\email{kotera@iap.fr, silk@iap.fr}

\begin{abstract}

 The recent 
detection of the gravitational wave source GW150914 by the LIGO collaboration motivates a speculative source for the origin of ultrahigh energy cosmic rays as a possible byproduct of the immense energies achieved in black hole mergers, provided that the black holes have spin as seems inevitable and there are  relic magnetic fields and disk debris remaining from the formation of the black holes or from their accretion history. We argue that given  the modest efficiency $< 0.01$ required per event per unit of gravitational wave energy release, merging black holes potentially provide an environment for accelerating cosmic rays to ultrahigh energies. The presence of tidally disrupted planetary or asteroidal debris could lead to associated Fast Radio Bursts.

\end{abstract}

\section{Introduction}\label{s:intro}

The extragalactic origin of ultrahigh energy cosmic rays remains a mystery, whereas galactic cosmic rays are generally interpreted as being Fermi accelerated via shocks around supernova remnants.  We point out here that the recent 
detection of the gravitational wave source GW150914 at a redshift $z\sim 0.09^{+0.03}_{-0.04}$ (luminosity distance $D_{\rm gw}=410^{+160}_{-180}\,$Mpc) by the
Laser Interferometer Gravitational Wave Observatory (LIGO) \citep{Abbott16} provides   new light on a speculative source for the origin of ultrahigh energy cosmic rays.  The inferred event is the merger of a black hole binary whose members have masses of $M_1=36^{+5}_{-4}M_\odot$ and $M_2=29^{+4}_{-4}M_\odot$, with final black hole mass $M=62^{+4}_{-4}M_\odot$ and $E_{\rm gw}=3.0^{+0.5}_{-0.5}M_\odot c^2\sim 5.4\times10^{54}\,\rm erg$ radiated in gravitational waves, and a peak gravitational wave luminosity of $L_{\rm gw,peak}=3.6^{+0.5}_{-0.4}\times 10^{56}\,\rm erg\,s^{-1}$. The inferred population rate is $\rho_{\rm BH}\sim 2-400\,\rm  Gpc^{-3}\,yr^{-1}$ \citep{Abbott16_rates}.

Could there be an electromagnetic counterpart? We speculate that the most likely counterparts are at extremely high energy where the injection requirements are modest in terms of mass. Here we focus on ultrahigh energy cosmic rays as a possible byproduct of the immense energies achieved in black hole (BH) mergers, especially if the black holes have spin as seems inevitable and there are any relic magnetic fields and debris remaining from the formation of the black holes.  The most likely long-lived debris would be a system of planets or asteroids that would be dynamically reenergized to produce an occasional plunging orbit in the course of the merger, with  tidal debris feeding an  accretion disk. We note that the  LIGO limit on the final spin amplitude of the
remnant BH  is $0.67^{+0.05}_{-0.07}$.
Ultrahigh energy neutrinos are another likely byproduct, produced by hadronic cosmic ray interactions with ambient debris, though no neutrino counterpart has been reported so far \citep{ANTARES16}. 

Our reasoning is driven (a) by the modest efficiency $\ltsim 0.03$ required per event per unit of gravitational  wave energy release, (b) by the frequency of such sources, which is  on the order of $\epsilon \sim 0.001$  of the number of core collapse supernovae \citep{Marchant16,Abbott16_rates}, estimated to occur at a rate $\sim 10^5\,\rm  Gpc^{-3}\,yr^{-1},$  (c) by the metal-poor and likely early-epoch  environment  of the massive star precursors whose associated chemical enrichment of the universe is also of order $\epsilon$ of the metal contribution from core collapse supernovae, namely $[Z] \sim 0.001,$ (d) by the iron-enriched nature of the residual debris around the merging black holes, and (e) by the transient nature of such extragalactic sources. These provide an intriguing basis for the hypothesis that we now explore, namely that merging black holes potentially provide an environment 
for accelerating cosmic rays to ultrahigh energies.

In Section~\ref{section:LEM} we demonstrate that BH mergers would have sufficient luminosity to power the acceleration of cosmic rays to the highest energies. In Section~\ref{section:transient}, we   argue that  these systems, as transient sources, can account for the total energy budget of the observed UHECRs  and for their distribution in the sky. Section~\ref{section:neutrinos} estimates the associated neutrino fluxes and shows that this model is compatible with the current IceCube neutrino sensitivities. We discuss possible sites for UHECR production within this scenario and a possible link with the signal reported by the {\it Fermi}-GBM in Section~\ref{section:discussion}.

\section{Electromagnetic luminosity}\label{section:LEM}

Although no specific literature can be found on the electromagnetic radiation counterparts from the merger of binary stellar BHs, studies of increasing numerical complexity predict such signatures for supermassive BH mergers. The simulations of gas and magnetic fields around the merging systems suggest that the motion of two BHs in a magnetically-dominated plasma could generate a magnetosphere and nebular structure similar to those inferred in pulsars, as well as collimated jets (e.g., \citealp{Milosavljevic05,ONeill09,Palenzuela09,Palenzuela10,Moesta10,Moesta12,Bode12,Giacomazzo12,Gold14}). The level of radiative flux generated is however uncertain, and subject to strong variabilities according to parameters and unknown structural details of the system. 

Most models are in line with the original Blandford-Znajek process \citep{Blandford77} that extracts the space-time rotational energy of the BHs to generate a powerful electromagnetic outflow. The same mechanisms can be applied to stellar BHs at the cost of rescaling the BH mass and the magnetic field. A rough estimate of the Poynting flux can then be derived \citep{Lyutikov11}
\begin{equation}\label{eq:LBZ}
L_{\rm BZ} = \frac{(GM)^3B^2}{c^5R} \sim  3.2\times 10^{46}\,{\rm erg\,s^{-1}}\,M_{100}^3B_{11}^2\frac{R_{\rm S}}{R}\ ,
\end{equation}
where $M$ is  the final black hole mass and $B=B_{11}/10^{11}\,{\rm G}$ is ­­ the strength of the external magnetic field. We have estimated the orbital radius $R$ as the Schwarzschild radius $R_{\rm S}=2GM/c^2\sim 3.0\times 10^7\,M_{100}\,$cm, with $M_{100}\equiv M/100\,M_\odot$. 

The magnetic field within the orbit is commonly estimated by assuming that a fraction $\eta_{\rm E}$ of the Eddington luminosity is tapped into magnetic luminosity, yielding values of $B\sim 3\times 10^6\,{\rm G}\,M_{100}^{-1/2}\eta_{\rm E}^{1/2}(R/R_{\rm S})^{-1}$ (e.g., \citealp{Lyutikov11}). Recent simulations demonstrate however that non-linear effects should amplify this field by up to  2 orders of magnitude \citep{Giacomazzo12}. One could also invoke an $\alpha\omega$-dynamo process as for pulsars and magnetars, that would generate fields of strength $B\sim 10^{12}\,{\rm G}\,(P/{300\,{\rm ms}})^{-1}$, with $P$ the spin period of the system \citep{Thompson93,Xu02}. The seed fields could be anchored to the remains of the accretion disk, the existence of which is proposed for example in \cite{Perna16}, that should rotate with period $P\sim 1-10\,$s, leading to a dynamo-generated field of $B\gtrsim 10^{10}\,$G. 

A stringent lower limit on the luminosity of any astrophysical outflow can be placed as a necessary condition to accelerate particles to energy $E$ \citep{LW09}: 
$L > 10^{45} (E/10^{20}\,{\rm eV})^2 Z^{-2}\,\rm erg\,s^{-1}$, with $Z$ the charge number of the particle. For a proton composition, this implies that the sources have to be exceptionally bright. Equation~(\ref{eq:LBZ}) suggests that a system like GW150914 should have sufficient power to accelerate particles up to the highest energies, as long as the magnetic field within the orbit can be $B\gtrsim 10^{11}\,$G.

\section{A transient candidate source for UHECRs}\label{section:transient}

Above $E > 10^{19}\,$eV, the observed cosmic-ray flux constrains the source population energy budget to ${E}_{\rm UHECR}\,\rho_{\rm 0}=10^{44.5}\,{\rm erg\,Mpc^{-3}\,yr^{-1}}$, requiring that each individual source  supplies an energy
\begin{equation}\label{Ebudget}
{E}_{\rm UHECR} \gtrsim 3.2\times 10^{53}\,{\rm erg}\,\left(\frac{\rho_{\rm 0}}{1\,{\rm Gpc^{-3}\,yr^{-1}}}\right)^{-1}\ , 
\end{equation}
with $\rho_0$ the source population rate at redshift $z=0$. This budget is not easily reached by most astrophysical populations. For BH mergers, the rates inferred by LIGO \citep{Abbott16_rates} imply $E_{\rm UHECR}\gtrsim 7.9\times10^{50}\,{\rm erg}\,(\rho_{\rm 0}/400\,{\rm Gpc^{-3}\,yr^{-1}})^{-1}$ and $E_{\rm UHECR}\gtrsim 1.6\times10^{53}\,{\rm erg}\,(\rho_{\rm 0}/2\,{\rm Gpc^{-3}\,yr^{-1}})^{-1}$, for the upper and lower rate limits respectively. Such energies represent a fraction of $<3\%$ of the energy released in gravitational waves by GW150914 ($E_{\rm gw}\sim 3.0\,M_\odot c^2\sim 5.4\times10^{54}\,{\rm erg\,s^{-1}}$). To achieve such energies, the system would be required to supply a luminosity $L_{\rm BZ}$ (Eq.~{\ref{eq:LBZ}) for time spans of 7 hours to 2 months. These durations constitute a comfortable fraction of the typical Blandford-Znajek timescale $t_{\rm BZ}=Mc^2/L_{\rm BZ}\sim 22\,M_{100}B_{11}^{-2}({R_{\rm S}}/{R})^2\,{\rm yr}$.  However, the Blandford-Znajek  process} would be maintained only as long as the black hole accretes after the merger. The relatively long disk accretion time needed by our model is best explained if the disk is sourced by tidal disruption of asteroids or planets. We note that the tidal radius for such a body of mass $m_{-18}\equiv m/10^{-18}\,M_\odot$ and size $r_{\rm km}\equiv r/ 1\,$km is about $r_{\rm t}\sim 4\times10^{11}\,{\rm cm}\,(M_{100}/m_{c})^{1/3}r_{\rm km}$. The orbital period for the debris is of order a day. Such disruptions are plausibly triggered by merger-perturbed orbits of residual asteroid clouds surrounding either or both of the merging black holes.

The absence of multiplets, namely cosmic ray events arriving with little angular separation in the sky, can be used to constrain the apparent number density of sources to $n_0>10^{-5}\,$Mpc$^{-3}$, even if particles are deflected to $\sim 30^\circ$ \citep{Abreu:2013kif}.
The low density of steady candidates: clusters of galaxies ($10^{-6}$\,Mpc$^{-3}$), FRI-type ($10^{-5}\,$Mpc$^{-3}$), and FRII-type radio-galaxies ($10^{-8}$\,Mpc$^{-3}$) is not compatible with these observations. For transient sources, on the other hand, the apparent $n_0$ and real $\rho_0$ number densities of proton UHECR sources are related via the cosmic ray arrival time spread $\delta t$ due to magnetic fields: $\rho_0\sim n_0/\delta t$ \citep{Murase_Takami09}. The time spread is of order $\delta t\sim 10^4\,$yrs for a $1^\circ$ deflection over 100\,Mpc, and even rare transient events (e.g., $\rho_0=1\,$Gpc$^{-3}$yr$^{-1}$) could mimic a rather dense population. The rates inferred by the LIGO observations for BH mergers are thus compatible with these observations. 

Note that the time delay due to the magnetic deflections will prevent us from observing UHECRs in correlation with the gravitational wave counterpart of a BH merger (this is valid for any transient source). The only direct evidence of an association between UHECRs and BH mergers can be obtained by the observation of gravitational waves in coincidence with high-energy neutrinos or FRBs, as discussed below. 

The statistically significant cosmic-ray excess above energy $5.7\times 10^{19}\,$eV reported by the Telescope Array (TA) within a $20^\circ$ radius circle centered at (${\rm R.A.} = 146.7$, ${\rm Dec.} = 43.2$) \citep{Abbasi14} can also be best accommodated with a transient source, due to the absence of powerful  source observed in the direction of this hotspot \citep{Renault-Tinacci16}. This BH merger scenario would be well-suited to account for this observation. 

The chemical composition of cosmic rays reported by the Auger Observatory is not compatible with a light composition at the highest energies \citep{Auger_icrc13,Auger_compo_2014a,Auger_compo_2014b}. The Telescope Array results seem to show the same trend within systematics \citep{TAicrc11,Pierog13,TA_auger_icrc13}. BH mergers stem from the core of massive stars and hence should be surrounded by metal-rich debris from before their collapse. These systems should thus offer a favorable site to produce and accelerate heavy nuclei.

\section{Associated neutrino fluxes}\label{section:neutrinos}

The secondary neutrino flux from UHECRs accelerated in an individual source at a distance $D_{\rm s}$ with luminosity $L_{\rm UHECR}$ can be estimated as 
\begin{eqnarray}
E_\nu^2\Phi_\nu&\sim&\frac{3}{8}f_\nu \, f_z \frac{L_{\rm UHECR}}{4\pi D_{\rm s}^2}\nonumber \\
&\sim& 3.7\times 10^{-7}\,{\rm GeV\,cm^{-2}\,s^{-1}}\,f_\nu \, f_z \times\nonumber\\
&&\frac{L_{\rm UHECR}}{10^{46.5}\,{\rm erg\,s^{-1}}}\left(\frac{D_{\rm s}}{410\,{\rm Mpc}}\right)^{-2}\ , 
\end{eqnarray}
with $f_z$ the redshift losses (comprised between 0.55 for a source at Hubble distance and 1 for a local source) and $f_\nu$ the optical depth to neutrino production. For the numerical estimate, as an upper bound to the neutrino flux, we have assumed that the luminosity derived in Eq.~(\ref{eq:LBZ}) is entirely tapped into UHECRs. This number is similar to the IceCube single source sensitivity, but is compatible with the absence of neutrino counterpart reported by {\sc Antares} and IceCube in the direction of GW150014 \citep{ANTARES16}, given the usually low value of $f_\nu$ in astrophysical sources. In the BH merger environment, the radiative and baryonic fields are not expected to be particularly intense. If particle acceleration happens in a jet-like structure created by Blandford-Znajek-type processes, the background fields should resemble those of standard gamma-ray burst scenarios. For instance for long gamma-ray bursts,  $f_\nu<10^{-4}$ \citep{Abbasi11_GRB,He12,Li12,Huemmer12}, and for short gamma-ray bursts the fields are expected to be of lower intensity. 
The experiments derive an upper bound on the neutrino energy from GW150014 of $10^{52-54}\,$erg. From the inferred energy budget of each source (see Section~\ref{section:transient}), one can already constrain the optical depth to neutrino production to $f_\nu < 1$. A more stringent evaluation of the rate of BH mergers would allow the derivation of a stronger upper bound on $f_\nu$. 

The diffuse all-flavor neutrino flux integrated over the entire source population reads 
$E_\nu^2\Phi_\nu=({3D_{\rm H}}/{32\pi})f_z\,f_\nu\,\rho_0\,E_{\rm UHECR} \sim 8.3\times 10^{-8} \,{\rm GeV\,cm^{-2}\,s^{-1}}\,f_\nu \, f_z$,
where $D_{\rm H}$ is the Hubble distance and the redshift loss factor $f_z\sim 0.55$ for a uniform source evolution history and $f_z\sim 2.5$  for an evolution following the star formation history as in \cite{Hopkins06} \citep{WB99,Fang14}. This estimate is comparable to the IceCube diffuse sensitivity limit (of order $10^{-7}\,{\rm GeV\,cm^{-2}\,s^{-1}\,sr^{-1}}$ at $\sim 10^{17}\,$eV) and the non-detection of neutrinos implies again an optical depth to neutrino production in the source of $f_\nu<1$.

\section{Discussion}\label{section:discussion}

We have argued that the production of UHECRs from a population of black hole mergers as observed in gravitational waves by LIGO can account, at first order, for all observational constraints of UHECRs (energy budget, global spectrum, arrival directions, composition and secondary messengers). Detailed simulations and models for acceleration sites would be needed in order to compute the exact composition of particles as a function of energy, and the associated spectral features due to energy losses on the radiative and baryonic backgrounds. 

It is however possible to infer these features from the pulsar model developed in \cite{Fang12,Fang13}, as it has been pointed out by \cite{Lyutikov11} that the magnetospheres of moving black holes resemble those of rotationally-powered pulsars, with pair formation fronts and outer gaps, and further out, the formation of a wind-driven cavity possibly surrounded by a nebular region. In this example, UHECR acceleration could happen either in the inner jet region as for gamma-ray bursts, or at the termination shock where the jet encounters the circumstellar medium, if the shock is strong enough, as in \cite{LKP15}. The environment of the pulsar could be similar to that described in \cite{Piro16} for neutron star mergers, with rather thin optical depths. 

One should note also that the present discussion can be applied to supermassive black hole mergers and lead to similar results, with the rescaling of the black hole mass and orbital magnetic field. 

We have argued in this work that, because of the transient nature of the source, it will not be possible to make a clear association between the (deflected and delayed) cosmic rays generated during merger and the emitted GW signal. The observation of correlated neutrino fluxes, that would confirm this model, also seems marginal given the current instrumental sensitivities and the thin source environment. A possible electromagnetic signature of this UHECR model could be Fast Radio Bursts, produced by Poynting flux-driven Alfven wings around circum-black hole asteroids. The presence of small bodies orbiting the black hole, that would be later disrupted to feed the accretion disk can indeed be invoked to maintain the accretion and thus the Blandford-Znajek process over a rather long timescale (see Section~\ref{section:transient}). The time-scale for disruption matches the inferred time-scale of the acceleration model and the tidal disruption radius is comparable to the range proposed to account for Fast Radio Bursts in the model of \cite{Mottez14}. Note that possible  asteroid-generated FRBs  could  precede the final merger by  typically longer  orbital time-scales, in addition to any similar events associated with the 
asteroid destruction on time-scales of order days.

The {\it Fermi} satellite Gamma-ray Burst Monitor reported the detection at $\sim 3\sigma$-level of a transient signal of luminosity $\sim 10^{49}\,{\rm erg\,s^{-1}}$ at photon energies $\sim 0.1-1$ MeV over 1 s that appeared 0.4 s after the gravitational wave signal. This signal was not detected by any other instrument and scientific discussion is ongoing (see e.g., \citealp{Savchenko16}).
If an external magnetic field of order $\gtrsim 5\times 10^{12}\,$G (as commonly observed in pulsars and magnetars) could be generated, Eq.~(\ref{eq:LBZ}) implies that the Blandford-Znajek process would extract enough electromagnetic luminosity to account for such an emission. 
Note that the Eddington luminosity, of order $L_{\rm Edd}\sim 1.4\times 10^{40}\,M_{100}\,{\rm erg\,s^{-1}}$, is much lower than the reported luminosity. 

The presence of debris clump of mass comparable to an asteroid, in the vicinity of the black hole, is again another possible mechanism for such an event: a similar phenomena, a slow GRB with an extremely faint optical counterpart has been suggested to be caused by disruption of a compact clump falling onto a neutron star \citep{Campana11}. 

We speculate that the early formation inferred for the BH precursors motivates a likely enhancement with overdense early-forming structures. These are likely to be in regions where the first stars may have formed, such as, for example, galactic bulges \citep{Belczynski10,howes15}.

\acknowledgments 
We thank E. Barausse for very fruitful discussions. KK acknowledges financial support from the PER-SU fellowship at Sorbonne Universit\'es and from the Labex ILP (reference ANR-10-LABX-63, ANR-11-IDEX-0004-02), JS was supported by ERC Project No. 267117 (DARK) hosted by Universit«e Pierre et
Marie Curie (UPMC) - Paris 6 and CEA Saclay, and also  acknowledges
the support of JHU by NSF grant OIA-1124403.


\begin{thebibliography}{0}
\expandafter\ifx\csname natexlab\endcsname\relax\def\natexlab#1{#1}\fi

\end{thebibliography}


\begin{thebibliography}{45}
\expandafter\ifx\csname natexlab\endcsname\relax\def\natexlab#1{#1}\fi

\bibitem[{{Aab} {et~al.}(2014{\natexlab{a}}){Aab}, {Abreu}, {Aglietta}, {Ahn},
  {Al Samarai}, {Albuquerque}, {Allekotte}, {Allen}, {Allison}, {Almela}, \&
  et~al.}]{Auger_compo_2014a}
{Aab}, A., {Abreu}, P., {Aglietta}, M., {Ahn}, E.~J., {Al Samarai}, I.,
  {Albuquerque}, I.~F.~M., {Allekotte}, I., {Allen}, J., {Allison}, P.,
  {Almela}, A., \& et~al. 2014{\natexlab{a}}, Phys. Rev. D, 90, 122005

\bibitem[{{Aab} {et~al.}(2014{\natexlab{b}}){Aab}, {Abreu}, {Aglietta}, {Ahn},
  {Al Samarai}, {Albuquerque}, {Allekotte}, {Allen}, {Allison}, {Almela}, \&
  et~al.}]{Auger_compo_2014b}
---. 2014{\natexlab{b}}, Phys. Rev. D, 90, 122006

\bibitem[{{Abbasi} {et~al.}(2011){Abbasi}, {Abdou}, {Abu-Zayyad}, {Adams},
  {Aguilar}, {Ahlers}, {Andeen}, {Auffenberg}, {Bai}, {Baker}, \&
  et~al.}]{Abbasi11_GRB}
{Abbasi}, R., {Abdou}, Y., {Abu-Zayyad}, T., {Adams}, J., {Aguilar}, J.~A.,
  {Ahlers}, M., {Andeen}, K., {Auffenberg}, J., {Bai}, X., {Baker}, M., \&
  et~al. 2011, Physical Review Letters, 106, 141101

\bibitem[{{Abbasi} {et~al.}(2014){Abbasi}, {Abe}, {Abu-Zayyad},
  {et~al.}}]{Abbasi14}
{Abbasi}, R.~U., {Abe}, M., {Abu-Zayyad}, T., {et~al.} 2014, ApJ Lett., 790,
  L21

\bibitem[{{Abbott} {et~al.}(2016{\natexlab{a}}){Abbott}, {Abbott}, {Abbott},
  {Abernathy}, {Acernese}, {Ackley}, {Adams}, {Adams}, {Addesso}, {Adhikari},
  \& et~al.}]{Abbott16}
{Abbott}, B.~P., {Abbott}, R., {Abbott}, T.~D., {Abernathy}, M.~R., {Acernese},
  F., {Ackley}, K., {Adams}, C., {Adams}, T., {Addesso}, P., {Adhikari}, R.~X.,
  \& et~al. 2016{\natexlab{a}}, Physical Review Letters, 116, 061102

\bibitem[{{Abbott} {et~al.}(2016{\natexlab{b}}){Abbott}, {Abbott}, {Abbott},
  {Abernathy}, {Acernese}, {Ackley}, {Adams}, {Adams}, {Addesso}, {Adhikari},
  \& et~al.}]{Abbott16_rates}
---. 2016{\natexlab{b}}, ArXiv e-prints: 1602.03842

\bibitem[{Abreu {et~al.}(2013)}]{Abreu:2013kif}
Abreu, P., {et~al.} 2013, JCAP, 1305, 009

\bibitem[{{Adri{\'a}n-Mart{\'{\i}}nez}
  {et~al.}(2016){Adri{\'a}n-Mart{\'{\i}}nez}, {Albert}, {Andr{\'e}}, {Anton},
  {Ardid}, {Aubert}, {Avgitas}, {Baret}, {Barrios-Mart{\'{\i}}}, \&
  et~al.}]{ANTARES16}
{Adri{\'a}n-Mart{\'{\i}}nez}, S., {Albert}, A., {Andr{\'e}}, M., {Anton}, G.,
  {Ardid}, M., {Aubert}, J.-J., {Avgitas}, T., {Baret}, B.,
  {Barrios-Mart{\'{\i}}}, J., \& et~al. 2016, ArXiv e-prints: 1602.05411

\bibitem[{{Belczynski} {et~al.}(2010){Belczynski}, {Dominik}, {Bulik},
  {O'Shaughnessy}, {Fryer}, \& {Holz}}]{Belczynski10}
{Belczynski}, K., {Dominik}, M., {Bulik}, T., {O'Shaughnessy}, R., {Fryer}, C.,
  \& {Holz}, D.~E. 2010, ApJ Lett., 715, L138

\bibitem[{{Blandford} \& {Znajek}(1977)}]{Blandford77}
{Blandford}, R.~D., \& {Znajek}, R.~L. 1977, MNRAS, 179, 433

\bibitem[{{Bode} {et~al.}(2012){Bode}, {Bogdanovi{\'c}}, {Haas}, {Healy},
  {Laguna}, \& {Shoemaker}}]{Bode12}
{Bode}, T., {Bogdanovi{\'c}}, T., {Haas}, R., {Healy}, J., {Laguna}, P., \&
  {Shoemaker}, D. 2012, ApJ, 744, 45

\bibitem[{{Campana} {et~al.}(2011){Campana}, {Lodato}, {D'Avanzo}, {Panagia},
  {Rossi}, {Della Valle}, {Tagliaferri}, {Antonelli}, {Covino}, {Ghirlanda},
  {Ghisellini}, {Melandri}, {Pian}, {Salvaterra}, {Cusumano}, {D'Elia},
  {Fugazza}, {Palazzi}, {Sbarufatti}, \& {Vergani}}]{Campana11}
{Campana}, S., {Lodato}, G., {D'Avanzo}, P., {Panagia}, N., {Rossi}, E.~M.,
  {Della Valle}, M., {Tagliaferri}, G., {Antonelli}, L.~A., {Covino}, S.,
  {Ghirlanda}, G., {Ghisellini}, G., {Melandri}, A., {Pian}, E., {Salvaterra},
  R., {Cusumano}, G., {D'Elia}, V., {Fugazza}, D., {Palazzi}, E., {Sbarufatti},
  B., \& {Vergani}, D.~S. 2011, Nature, 480, 69

\bibitem[{{Fang} {et~al.}(2014){Fang}, {Kotera}, {Murase}, \&
  {Olinto}}]{Fang14}
{Fang}, K., {Kotera}, K., {Murase}, K., \& {Olinto}, A.~V. 2014, Phys Rev D,
  90, 103005

\bibitem[{{Fang} {et~al.}(2012){Fang}, {Kotera}, \& {Olinto}}]{Fang12}
{Fang}, K., {Kotera}, K., \& {Olinto}, A.~V. 2012, ApJ, 750, 118

\bibitem[{{Fang} {et~al.}(2013){Fang}, {Kotera}, \& {Olinto}}]{Fang13}
---. 2013, J. Cos. and Astro. Phys., 3, 10

\bibitem[{{Giacomazzo} {et~al.}(2012){Giacomazzo}, {Baker}, {Miller},
  {Reynolds}, \& {van Meter}}]{Giacomazzo12}
{Giacomazzo}, B., {Baker}, J.~G., {Miller}, M.~C., {Reynolds}, C.~S., \& {van
  Meter}, J.~R. 2012, ApJ Lett., 752, L15

\bibitem[{{Gold} {et~al.}(2014){Gold}, {Paschalidis}, {Etienne}, {Shapiro}, \&
  {Pfeiffer}}]{Gold14}
{Gold}, R., {Paschalidis}, V., {Etienne}, Z.~B., {Shapiro}, S.~L., \&
  {Pfeiffer}, H.~P. 2014, Phys. Rev. D, 89, 064060

\bibitem[{{He} {et~al.}(2012){He}, {Liu}, {Wang}, {Nagataki}, {Murase}, \&
  {Dai}}]{He12}
{He}, H.-N., {Liu}, R.-Y., {Wang}, X.-Y., {Nagataki}, S., {Murase}, K., \&
  {Dai}, Z.-G. 2012, ApJ, 752, 29

\bibitem[{{Hopkins} \& {Beacom}(2006)}]{Hopkins06}
{Hopkins}, A.~M., \& {Beacom}, J.~F. 2006, ApJ, 651, 142

\bibitem[{{Howes} {et~al.}(2015){Howes}, {Casey}, {Asplund}, {Keller}, {Yong},
  {Nataf}, {Poleski}, {Lind}, {Kobayashi}, {Owen}, {Ness}, {Bessell}, {da
  Costa}, {Schmidt}, {Tisserand}, {Udalski}, {Szyma{\'n}ski}, {Soszy{\'n}ski},
  {Pietrzy{\'n}ski}, {Ulaczyk}, {Wyrzykowski}, {Pietrukowicz}, {Skowron},
  {Koz{\l}owski}, \& {Mr{\'o}z}}]{howes15}
{Howes}, L.~M., {Casey}, A.~R., {Asplund}, M., {Keller}, S.~C., {Yong}, D.,
  {Nataf}, D.~M., {Poleski}, R., {Lind}, K., {Kobayashi}, C., {Owen}, C.~I.,
  {Ness}, M., {Bessell}, M.~S., {da Costa}, G.~S., {Schmidt}, B.~P.,
  {Tisserand}, P., {Udalski}, A., {Szyma{\'n}ski}, M.~K., {Soszy{\'n}ski}, I.,
  {Pietrzy{\'n}ski}, G., {Ulaczyk}, K., {Wyrzykowski}, {\L}., {Pietrukowicz},
  P., {Skowron}, J., {Koz{\l}owski}, S., \& {Mr{\'o}z}, P. 2015, Nature, 527,
  484

\bibitem[{{H{\"u}mmer} {et~al.}(2012){H{\"u}mmer}, {Baerwald}, \&
  {Winter}}]{Huemmer12}
{H{\"u}mmer}, S., {Baerwald}, P., \& {Winter}, W. 2012, Physical Review
  Letters, 108, 231101

\bibitem[{{Lemoine} {et~al.}(2015){Lemoine}, {Kotera}, \& {P{\'e}tri}}]{LKP15}
{Lemoine}, M., {Kotera}, K., \& {P{\'e}tri}, J. 2015, JCAP, 7, 016

\bibitem[{Lemoine \& Waxman(2009)}]{LW09}
Lemoine, M., \& Waxman, E. 2009, JCAP, 0911, 009

\bibitem[{{Li}(2012)}]{Li12}
{Li}, Z. 2012, Phys. Rev. D, 85, 027301

\bibitem[{{Lyutikov}(2011)}]{Lyutikov11}
{Lyutikov}, M. 2011, Phys Rev D, 83, 124035

\bibitem[{{Marchant} {et~al.}(2016){Marchant}, {Langer}, {Podsiadlowski},
  {Tauris}, \& {Moriya}}]{Marchant16}
{Marchant}, P., {Langer}, N., {Podsiadlowski}, P., {Tauris}, T., \& {Moriya},
  T. 2016, ArXiv e-prints: 1601.03718

\bibitem[{{Milosavljevi{\'c}} \& {Phinney}(2005)}]{Milosavljevic05}
{Milosavljevi{\'c}}, M., \& {Phinney}, E.~S. 2005, ApJ Lett., 622, L93

\bibitem[{{Moesta} {et~al.}(2012){Moesta}, {Alic}, {Rezzolla}, {Zanotti}, \&
  {Palenzuela}}]{Moesta12}
{Moesta}, P., {Alic}, D., {Rezzolla}, L., {Zanotti}, O., \& {Palenzuela}, C.
  2012, ApJ Lett., 749, L32

\bibitem[{{Moesta} {et~al.}(2010){Moesta}, {Palenzuela}, {Rezzolla}, {Lehner},
  {Yoshida}, \& {Pollney}}]{Moesta10}
{Moesta}, P., {Palenzuela}, C., {Rezzolla}, L., {Lehner}, L., {Yoshida}, S., \&
  {Pollney}, D. 2010, Phys Rev D, 81, 064017

\bibitem[{{Mottez} \& {Zarka}(2014)}]{Mottez14}
{Mottez}, F., \& {Zarka}, P. 2014, A\&A, 569, A86

\bibitem[{{Murase} \& {Takami}(2009)}]{Murase_Takami09}
{Murase}, K., \& {Takami}, H. 2009, ApJ Letters, 690, L14

\bibitem[{{O'Neill} {et~al.}(2009){O'Neill}, {Miller}, {Bogdanovi{\'c}},
  {Reynolds}, \& {Schnittman}}]{ONeill09}
{O'Neill}, S.~M., {Miller}, M.~C., {Bogdanovi{\'c}}, T., {Reynolds}, C.~S., \&
  {Schnittman}, J.~D. 2009, ApJ, 700, 859

\bibitem[{{Palenzuela} {et~al.}(2009){Palenzuela}, {Anderson}, {Lehner},
  {Liebling}, \& {Neilsen}}]{Palenzuela09}
{Palenzuela}, C., {Anderson}, M., {Lehner}, L., {Liebling}, S.~L., \&
  {Neilsen}, D. 2009, Physical Review Letters, 103, 081101

\bibitem[{{Palenzuela} {et~al.}(2010){Palenzuela}, {Lehner}, \&
  {Liebling}}]{Palenzuela10}
{Palenzuela}, C., {Lehner}, L., \& {Liebling}, S.~L. 2010, Science, 329, 927

\bibitem[{{Perna} {et~al.}(2016){Perna}, {Lazzati}, \& {Giacomazzo}}]{Perna16}
{Perna}, R., {Lazzati}, D., \& {Giacomazzo}, B. 2016, ArXiv e-prints:
  1602.05140

\bibitem[{{Pierog}(2013)}]{Pierog13}
{Pierog}, T. 2013, 10

\bibitem[{{Piro} \& {Kollmeier}(2016)}]{Piro16}
{Piro}, A.~L., \& {Kollmeier}, J.~A. 2016, ArXiv e-prints:1601.02625

\bibitem[{Renault-Tinacci {et~al.}(2016)Renault-Tinacci, {Kotera}, \&
  {Olinto}}]{Renault-Tinacci16}
Renault-Tinacci, N., {Kotera}, K., \& {Olinto}, A.~V. 2016, in prep.

\bibitem[{{Savchenko} {et~al.}(2016){Savchenko}, {Ferrigno}, {Mereghetti},
  {Natalucci}, {Bazzano}, {Bozzo}, {Courvoisier}, {Brandt}, {Hanlon},
  {Kuulkers}, {Laurent}, {Lebrun}, {Roques}, {Ubertini}, \&
  {Weidenspointner}}]{Savchenko16}
{Savchenko}, V., {Ferrigno}, C., {Mereghetti}, S., {Natalucci}, L., {Bazzano},
  A., {Bozzo}, E., {Courvoisier}, T.~J.-L., {Brandt}, S., {Hanlon}, L.,
  {Kuulkers}, E., {Laurent}, P., {Lebrun}, F., {Roques}, J.~P., {Ubertini}, P.,
  \& {Weidenspointner}, G. 2016, ArXiv e-prints: 1602.04180

\bibitem[{{Tameda} {et~al.}(2011)}]{TAicrc11}
{Tameda}, Y., {et~al.} 2011, 32nd International Cosmic Ray Conference, Beijing,
  China, August 2011

\bibitem[{{Telescope Array} {et~al.}(2013){Telescope Array}, {Pierre Auger
  Collaborations}, {:}, {Abu-Zayyad}, {Allen}, {Anderson}, {Azuma},
  {Barcikowski}, {Belz}, {Bergman}, \& et~al.}]{TA_auger_icrc13}
{Telescope Array}, T., {Pierre Auger Collaborations}, {:}, {Abu-Zayyad}, T.,
  {Allen}, M., {Anderson}, R., {Azuma}, R., {Barcikowski}, E., {Belz}, J.~W.,
  {Bergman}, D.~R., \& et~al. 2013, ArXiv e-prints

\bibitem[{{The Pierre Auger Collaboration} {et~al.}(2013){The Pierre Auger
  Collaboration}, {Aab}, {Abreu}, {Aglietta}, {Ahlers}, {Ahn}, {Albuquerque},
  {Allekotte}, {Allen}, {Allison}, \& et~al.}]{Auger_icrc13}
{The Pierre Auger Collaboration}, {Aab}, A., {Abreu}, P., {Aglietta}, M.,
  {Ahlers}, M., {Ahn}, E.-J., {Albuquerque}, I., {Allekotte}, I., {Allen}, J.,
  {Allison}, P., \& et~al. 2013, ArXiv e-prints: 1307.5059

\bibitem[{{Thompson} \& {Duncan}(1993)}]{Thompson93}
{Thompson}, C., \& {Duncan}, R.~C. 1993, ApJ, 408, 194

\bibitem[{{Waxman} \& {Bahcall}(1999)}]{WB99}
{Waxman}, E., \& {Bahcall}, J. 1999, Phys. Rev. D, 59, 023002

\bibitem[{{Xu} {et~al.}(2002){Xu}, {Wang}, \& {Qiao}}]{Xu02}
{Xu}, R., {Wang}, H., \& {Qiao}, G. 2002, Chinese Journal of Astronomy \&
  Astrophysics, 2, 533

\end{thebibliography}
\end{document}